\documentclass{ws-procs975x65}

\begin{document}

\title{New results at 3PN via an effective field theory of gravity}

\author{Rafael A. Porto}

\address{Department of Physics, Carnegie Mellon University, Pittsburgh, PA 15213}

\begin{abstract}
NRGR, an Effective Field Theory approach to gravity, has emerged
as a powerful tool to systematically compute higher order
corrections in the Post-Newtonian expansion. Here we discuss in
somehow more detail the recently reported new results for the
spin-spin gravitational potential at third Post-Newtonian order.
\end{abstract}

\bodymatter


\section*{}

A new approach, coined NRGR, has been recently introduced as a new technique
to systematically calculate within the Post-Newtonian expansion
via an effective field theory approach \cite{nrgr,nrgr2,nrgr3}.
The purpose of this contribution is to elaborate upon the new results recently reported
\cite{eih} for the spin-spin potential.
Further details will appear in a forthcoming publication.\\
The extension of NRGR to include spin effects\cite{nrgr3} can be
achieved by adding rotational degrees of freedom ($e^\mu_I$) in
the worldline action. The generalized angular velocity is given by
$\Omega^{\mu\nu}=e^{\nu J}\frac{De^\mu_J}{d\lambda}$, and the spin
$S_{\mu\nu}$ is introduced as the conjugate momentum. The form of
the world-line action is then fixed by reparameterization
invariance,
\begin{equation} \label{action}
S=-\sum_q \left(\int p^\mu_q u^q_{\mu}d\lambda_q + \int
\frac{1}{2}S_q^{\mu\nu}\Omega^q_{\mu\nu} d\lambda_q\right),
\end{equation}
where $\lambda_q$ is the proper length for the $q$'th worldline.
The Papapetrou equations follow from (\ref{action}) \cite{nrgr3}.
Higher dimensional terms describing finite size effects have been
left out although its inclusion is straightforward
\cite{nrgr,nrgr3,eih}. In order to account for the correct number
of degrees of freedom a so called spin supplementarity conditions
(SSC) is added to the equations of motion (EOM). The most
convenient choices are the covariant, $S^{\mu\nu} p_\nu=0$, and
Newton-Wigner (NW), $S^{\mu\nu}p_{\nu} = m S^{\mu 0}$, SSC. Notice
that the latter is not covariant, however it can be shown to have
the advantage that the algebra reduces to a canonical structure
(up to subleading corrections) after Dirac brackets are imposed
\cite{regge}. The leading order spin-spin and spin-orbit effects
were shown to follow from the potentials within the NW SSC
\cite{nrgr3}. The 3PN spin-spin potential, $V^{ss}_{3pn}$, was
recently obtained \cite{eih}, so that the spin-spin part of the
EOM followed by means of the traditional Hamilton-Lagrange
approach. As we shall see this is a correct statement up to 4PN
where curvature effects
in the algebra start to play a role.\\

The spin-gravity coupling in (\ref{action}) can be rewritten by
introducing the spin coefficients, $\omega^{ab}_\mu$, as
\begin{equation} \label{act2}
S_{spin} \sim  -\frac{1}{2} \int S_{Lab}\omega^{ab}_\mu u^\mu
d\lambda,
\end{equation}
with $S_L^{ab}$ the spin tensor in a local Lorentzian frame
defined by the vierbein $e^a_\mu$. In this basis the co-rotating
frame is given by $e^J_\mu = \Lambda^J_a(\tau)e^a_\mu$ with
$\Lambda$ a Lorentz boost. By further expanding (\ref{act2}) in
the weak gravity limit one obtains the Feynman rules
\cite{nrgr3,eih}. Let us emphasize here that the spin tensor
appearing in the vertex rules is the one defined in the local
frame, where the NW SSC was chosen \footnote{One could chose to
expand the action in terms of $S^{\mu\nu}$. However, to obtain the
EOM from the potentials one would need to account for a more
complicated spin algebra.}. Before imposing the SSC one can show
that the algebra for the phase space variables
$(x^\mu,p^\nu,S^{ab}_L)$ is given by
\begin{eqnarray}
\{x^\mu ,{\cal P}_\alpha\}&=& \delta^\mu_\alpha,\;\;\; \{x^\mu,p_\alpha\}=\delta^\mu_\alpha  \\
\{{\cal P}^\alpha,{\cal P}^\beta\}&=& 0, \;\;\; \{x^\mu,x^\nu\}= 0, \;\;\; \{p^\alpha,p^\beta\} = \frac{1}{2}{R^{\alpha\beta}}_{ab}S_L^{ab}, \label{pp}\\
\{x^\mu ,S^{ab}_L\}&=& 0, \;\;\; \{p^\alpha,S^{ab}_L\}= 0, \;\;\; \{{\cal P}^\alpha,S^{ab}_L\}= 0 \\
\{S^{ab}_L,S^{cd}_L\} &=& \eta^{ac} S_L^{bd}
+\eta^{bd}S_L^{ac}-\eta^{ad} S_L^{bc}-\eta^{bc}
S_L^{ad},\label{als}
\end{eqnarray}
where $p^\mu$ is related to the canonical momentum by $ {\cal
P}^\mu = p^\mu - \frac{1}{2}\omega^\mu_{ab} S_L^{ab}$. After the
SSC is enforced a Dirac structure emerges. In flat space-time the
NW SSC will preserve the canonical structure in the reduced space
$(x^i,p^i,S_L^i)$, with $S_L^i=\epsilon^{ijk}S_L^{jk}$ the spin
three vector. In a curved background however, the algebra turns
out to be
\begin{eqnarray}
\{x^i,{\cal P}_j\}&=& \delta^i_j + ...\\
\{x^i,x^j\}&=& 0 + ...\\
\{{\cal P}^i,{\cal P}^j\} &=& 0 + ... \label{pp2n}\\
\{x^i,S_L^i\}&=& 0 + ...\\
\{{\cal P}^j,S_L^i\}&=& 0 + ...\\
\{S_L^i,S_L^j\} &=& \epsilon^{ijk}S_L^k\label{als2},
\end{eqnarray}
with the ellipses representing a series of
``curvature$\times$spin" terms\footnote{For example, in the
electromagnetic case \cite{regge}, similar to ours after the
identification $A_\mu \sim \omega^{ab}_\mu S_{ab}$, the Dirac
structure (in the covariant SSC) turns out to be a very cumbersome
expression.}. In principle we should worry about these curvature
effects, however we will show by standard power counting, its
effects in the spin-spin EOM are subleading and the canonical
procedure is accurate up to 4PN. The reason is somehow intuitive.
To get a correction coming from the algebra to the $\vec{S}_1\cdot
\vec{S}_2$ piece of the EOM for particle 1, one needs to consider
the $\vec{S}_2$ part of the spin-orbit Hamiltonian. The latter
scales as $v^3$ relatively to the Newtonian term. We know on the
other hand that the spin-orbit EOM does not receive any
corrections at leading order (1.5PN). This is a not trivial
statement given the fact that it could be modified by a non
trivial commutator with the leading order Hamiltonian. Therefore,
``algebra corrections" should start at 2.5PN. To get a correction
to the spin-spin EOM we would then need to hook up a 1.5PN
spin-orbit Hamiltonian with a 2.5PN algebra term, effectively a
4PN correction. Let us consider for instance the commutator
$\{x^i,x^j\}$ as an example. This commutator in the NW SSC will
receive corrections scaling as (schematically) $\sim R
x^2\frac{S}{m^2}+...$, with $R$ the Riemann tensor. On the other
hand, in the covariant SSC, this bracket is modified \cite{regge}
to $\{x^i,x^j\} = \frac{S^{ji}}{m^2}$, whose net effect in the EOM
is a 1.5PN term, necessary indeed to prove the equivalence for
different SSCs \cite{nrgr3}. The new term has now an extra factor
scaling as $\partial^2 h_{00} x^2$ at leading order ($R \sim
\partial^2 h_{00}$). In the weak gravity approximation, $h_{00}
\sim v^2$, so that the algebra-correction effectively starts at
2.5PN as we had foreseen\footnote{Other corrections could go as $R
\frac{S^2}{m^2}\frac{S}{m^2}$ and can be shown to be subleading.}.

Let us add a few words on the NW SSC in a curved background and
the spin choice. The NW condition implies (for each particle)
\begin{equation}
mS_L^{a0}=S_L^{ab}p_b \to S_{L(pn)}^{i0}=\frac{1}{2}v^jS_{L(pn)}^{ij}+O(v^4)
\end{equation}
where $S_{pn}^{ij}$ is the spin tensor in the original PN frame
($S_L^{ab} = e^a_\mu e^b_\nu S_{pn}^{\mu\nu}$), and $v^i$ the
three coordinate velocity\footnote{Depending on the frame choice
the $O(v^4)$ piece will change, however, the leading order
condition stays the same regardless of the choice.}. One can also
relate both spin tensors (we removed the {\it pn} label for
simplicity),
\begin{equation}
S^{ij}_{1L} = S^{ij}_1 + S_1^{ik}h^j_k - S_1^{jk}h^i_k + ... \sim S_1^{ij} + 4 \frac{G_Nm_2}{r}S_1^{ij}+...\label{spn}
\end{equation}
and then transform the EOM in terms of $S^{i}$, and hence to the
covariant SSC.

As we said above spin-spin subleading effects can be computed
regardless of algebra corrections up to 4PN. This is however not
true for subleading spin-orbit effects at 2.5PN \cite{owen,buon},
where these corrections start to contribute. We will thus finish
this short contribution with yet another approach which will
naturally overcome these difficulties in a more natural fashion.

Going back to the covariant SSC it is easy to show, from Papapetrou equations,
\begin{equation}
p^{\alpha}= m
u^{\alpha}-\frac{1}{2m}R_{\beta\nu\rho\sigma}S^{\alpha\beta}S^{\rho\sigma}u^{\nu}\label{up}.
\end{equation}

Notice that $p \cdot u = m$ {\it on shell} (once the SSC is
obeyed). One can thus show that the action (\ref{action}) is
equivalent to the following Routhian,
\begin{equation} \label{act}
{\cal R} = -\sum_q \left(\int m_q \sqrt{u^2_q}d\lambda_q + \int
\frac{1}{2}S_{Lq}^{ab}\omega_{ab\mu} u^\mu_q  - \frac{1}{2m_q}R_{d
e a b}(x_q)S^{c d}_{Lq} S^{a b}_{Lq} u^e_q u^q_c
~d\lambda_q\right).
\end{equation}
There is an extra piece, $S^{ab}_L S_{Lab}$, not shown. This term
does not affect the spin EOM since it is a Casimir operator.
However, it enters in the worldline evolution in the form of a
spin dependent mass. The EOM are,
\begin{equation}
\frac{\delta {\cal R}}{\delta x^\mu}=0, \;\;\; \frac{d
S^{ab}_L}{d\tau} = \{S^{ab}_L,{\cal R}\}\label{eom},
\end{equation}
which can be shown to reproduce eq. (\ref{up}) and Papapetrou
equations on shell, e.g. on the constraint surface
$S^{ab}_Lp_b=0$\footnote{A similar Routhian was advocated in Ref.
\cite{yb} with $S^{ab}_Lu_b$ as SSC. A Routhian was also shown to
be very convenient in Ref. \cite{sch}}. To obtain Post-Newtonian
corrections one calculates ${\cal R}$ perturbatively. Notice that,
had we imposed the SSC in (\ref{act}) one would get rid of the
Riemann term and end up in an approach equivalent to what we
discussed before. We will proceed in a different way and we will
impose the SSC condition after the EOM for ($x^i, S^{ij}_L$) are
obtained from (\ref{eom}), while keeping the power counting rules
for spin as before \cite{nrgr3}, e.g. $S^{0k}_L \sim v^i
S^{ik}_L$. The advantage of this approach is that one does not
have to worry about complicated algebraic structures. The price to
pay is the need of a spin tensor rather than a vector. As an
example let us compute the leading order spin-orbit contribution
to the spin EOM \footnote{The leading spin-spin EOM does not
include $S^{a0}$ and thus follows the exact same steps.}. The
spin-orbit potential is given by (we dropped $L$ for simplicity)
\begin{equation}
V^{so}_{1.5pn} =
\frac{G_Nm_2}{r^2}n^j\left(S^{j0}_1+S^{jk}_1(v_1^k-2v^k_2)\right)
+ 1 \leftrightarrow 2,
\end{equation}
with $n^j=(x_1-x_2)^j$. The relevant piece of the algebra is the
commutator
\begin{equation}
\{S^i,S^{j0}\} = \epsilon^{ijk}S^{0k} = v^iS^j-v^jS^i + ...,
\end{equation}
which follows from (\ref{als}) in the covariant SSC. Using
(\ref{eom}) one gets,
\begin{equation}
\frac{d\vec{S}_1}{dt} = 2\left(1+\frac{m_2}{m_1}\right)\frac{\mu
G_N}{r^2}(\vec{n}\times\vec{v})\times\vec{S}_1 -
\frac{m_2G_N}{r^2}(\vec{S}_1\times\vec{n})\times \vec{v_1}
\end{equation}
with $\mu$ the reduced mass and $\vec{v}$ the relative velocity. This agrees with the known result after the shift
\cite{nrgr3},
\begin{equation}
\vec{S}_1 \to (1-\frac{1}{2}\vec{v}_1^2)\vec{S}_1 + \frac{1}{2}\vec{v}_1(\vec{v}_1\cdot\vec{S}_1).
\end{equation}

Details and higher order computations will appear in a forthcoming
paper.\\ We would like to thank Gerhard Sch\"afer for helpful
discussions and bringing to our attention the subtleties of the
algebraic approach. We thank Ira Rothstein for helpful comments
and collaboration. This work was supported by DOE contracts
DOE-ER-40682-143 and DEAC02-6CH03000.

\end{document}